\documentclass[aps,prl,reprint,superscriptaddress,showpacs,floatfix]{revtex4-1}

\usepackage{graphicx}
\usepackage{amssymb} 
\usepackage{dcolumn}
\usepackage{bm}
\usepackage[mathlines]{lineno}

\makeatletter
\renewcommand*{\@fnsymbol}[1]{\ensuremath{\ifcase#1\or \dagger\or *\or \ddagger\or
\mathsection\or \mathparagraph\or \|\or **\or \dagger\dagger \or
\ddagger\ddagger \else\@ctrerr\fi}} \makeatother

\usepackage{soul,color}
\usepackage[colorlinks,linkcolor=blue,anchorcolor=blue,citecolor=blue,urlcolor=blue]{hyperref}

\begin{document}
\title{Towards room-temperature superconductivity
       in low-dimensional C$_{60}$ nanoarrays:\\
       An {\em ab initio} study}



\author{Dogan Erbahar}
\thanks{These two authors contributed equally.}
\affiliation{Physics and Astronomy Department,
             Michigan State University,
             East Lansing, Michigan 48824, USA}
\affiliation{Physics Department,
             Gebze Technical University,
             41400, Kocaeli, Turkey}

\author{Dan Liu}
\thanks{These two authors contributed equally.}
\affiliation{Physics and Astronomy Department,
             Michigan State University,
             East Lansing, Michigan 48824, USA}

\author{Savas Berber}
\affiliation{Physics Department,
             Gebze Technical University,
             41400, Kocaeli, Turkey}

\author{David Tom\'anek}
\email[E-mail: ]{tomanek@msu.edu}
\affiliation{Physics and Astronomy Department,
             Michigan State University,
             East Lansing, Michigan 48824, USA}

\date{\today }
\begin{abstract}
We propose to raise the critical temperature $T_c$ for
superconductivity in doped C$_{60}$ molecular crystals by
increasing the electronic density of states at the Fermi level
$N(E_F)$ and thus the electron-phonon coupling constant in
low-dimensional C$_{60}$ nanoarrays. We consider both electron and
hole doping and present numerical results for $N(E_F)$, which
increases with decreasing bandwidth of the partly filled $h_u$ and
$t_{1u}$ derived frontier bands with decreasing coordination
number of C$_{60}$. Whereas a significant increase of $N(E_F)$
occurs in 2D arrays of doped C$_{60}$ intercalated in-between
graphene layers, we propose that the highest $T_c$ values
approaching room temperature may occur in bundles of nanotubes
filled by 1D arrays of externally doped C$_{60}$ or La@C$_{60}$,
or in diluted 3D crystals, where quasi-1D arrangements of C$_{60}$
form percolation paths.
\end{abstract}


\pacs{
81.05.ub, 
73.22.-f, 
74.70.Wz,  
71.15.Mb 
 }


\maketitle


%
%
%
%
%

The quest for room-temperature superconductivity has lost nothing
of its appeal during the 30-year long intense search following the
observation of superconductivity in cuprate perovskites, with the
critical temperature $T_c$ rising from the 30-K range in the
La-Ba-Cu-O system~\cite{LBCO1986} to 77~K in
Y-Ba-Cu-O~\cite{YBCO1987}. Current record $T_c$ values of
$133$~K in the doped HgBa$_2$Ca$_2$Cu$_3$O$_8$
perovskite~\cite{Highest-TC} and $203$~K in sulfur
hydride~\cite{drozdov2015} have only been observed under high
pressure. Progress in raising $T_c$ significantly further has
lagged behind expectations.
Whereas the microscopic origin of superconductivity is still being
speculated about in high-$T_c$ compounds, the rather high $T_c$
values observed in doped solid C$_{60}$, possibly even exceeding
$60$~K~\cite{C60SC70K}, result from strong electron-phonon
coupling caused by the dynamical Jahn-Teller effect on individual
fullerene molecules~\cite{{DT053},{DT057}}. In alkali-doped
M$_3$C$_{60}$ molecular solids, $T_c$ could be quantitatively
reproduced~\cite{{DT053},{DT057}} using the McMillan
equation~\cite{McMillan68}. The key behind a substantial
electron-phonon coupling constant is one of its factors, namely a
high
density of states (DOS) at the Fermi level $N(E_F)$, which depends
on the particular element M used to intercalate bulk C$_{60}$.

Here we propose a way to further increase $T_c$ for
superconductivity by increasing $N(E_F)$ and thus the
electron-phonon coupling constant $\lambda$ by reducing the
C$_{60}$ coordination number $Z$ in doped low-dimensional C$_{60}$
nanoarrays. We considered intercalation by both electron donors
and acceptors, as well as electron doping in a solid formed of
La@C$_{60}$ endohedral complexes. We found that $N(E_F)$ increases
with decreasing bandwidth of the partly filled $h_u$ HOMO and
$t_{1u}$ LUMO derived frontier bands, which may be achieved by
reducing the coordination number of C$_{60}$. Whereas $N(E_F)$
increases significantly by changing from 3D C$_{60}$ crystals to
2D arrays of doped fullerenes intercalated in-between graphene
layers, $N(E_F)$ reaches its maximum in doped quasi-1D arrays of
C$_{60}$ molecules inside $(10,10)$ carbon nanotubes (CNTs),
forming C$_{60}$@CNT peapods. Whereas partial filling of the
$t_{1u}$-derived band of C$_{60}$ may be achieved by adsorbing K
atoms on the peapod surface, the desired depopulation of the
$h_u$-derived band by adsorbed F is not possible. Our results
indicate that the highest $T_c$ value approaching room temperature
may occur in electron-doped C$_{60}$ peapod arrays or in diluted
3D crystals, where quasi-1D arrangements of C$_{60}$ form
percolation paths.


%

We performed density functional theory (DFT) calculations to
obtain insight into the effect of geometrical arrangement of
fullerenes on the electronic structure of C$_{60}$ intercalation
compounds. We used the Perdew-Zunger~\cite{Perdew81} form of the
spin-polarized exchange-correlation functional in the local
density approximation to DFT, as implemented in the
\textsc{SIESTA} code~\cite{soler-tsmfaioms2002}, which correctly
reproduces the inter-layer spacing and interaction in graphitic
structures. The valence electrons were described by
norm-conserving Troullier-Martins
pseudopotentials~\cite{Troullier91} with partial core corrections
in the Kleinman-Bylander factorized form~\cite{Kleinman82}. We
used a double-zeta polarized basis and limited the range of the
localized orbitals in such a way that the energy shift caused by
their spatial confinement was no more than
10~meV~\cite{SIESTA_PAO}. The Brillouin zone of a 3D lattice of
C$_{60}$ molecules was sampled by $10{\times}10{\times}10$
k-points, that of a 2D lattice by $10{\times}10$ k-points, and
that of decoupled 1D chains of C$_{60}$ molecules inside a
nanotube by $10$ k-points. The DOS was convoluted by
$0.02$~eV$^{-1}$. In a periodic arrangement, 1D structures were
separated by $15$~{\AA} thick vacuum regions and 2D structures by
$13$~{\AA} thick vacuum regions. The charge density and the
potentials were determined on a real-space grid with a mesh cutoff
energy of $180$~Ry, which was sufficient to achieve a total energy
convergence of better than 2~meV/atom.

\begin{figure}[t]
\begin{center}
\includegraphics[width=0.90\columnwidth]{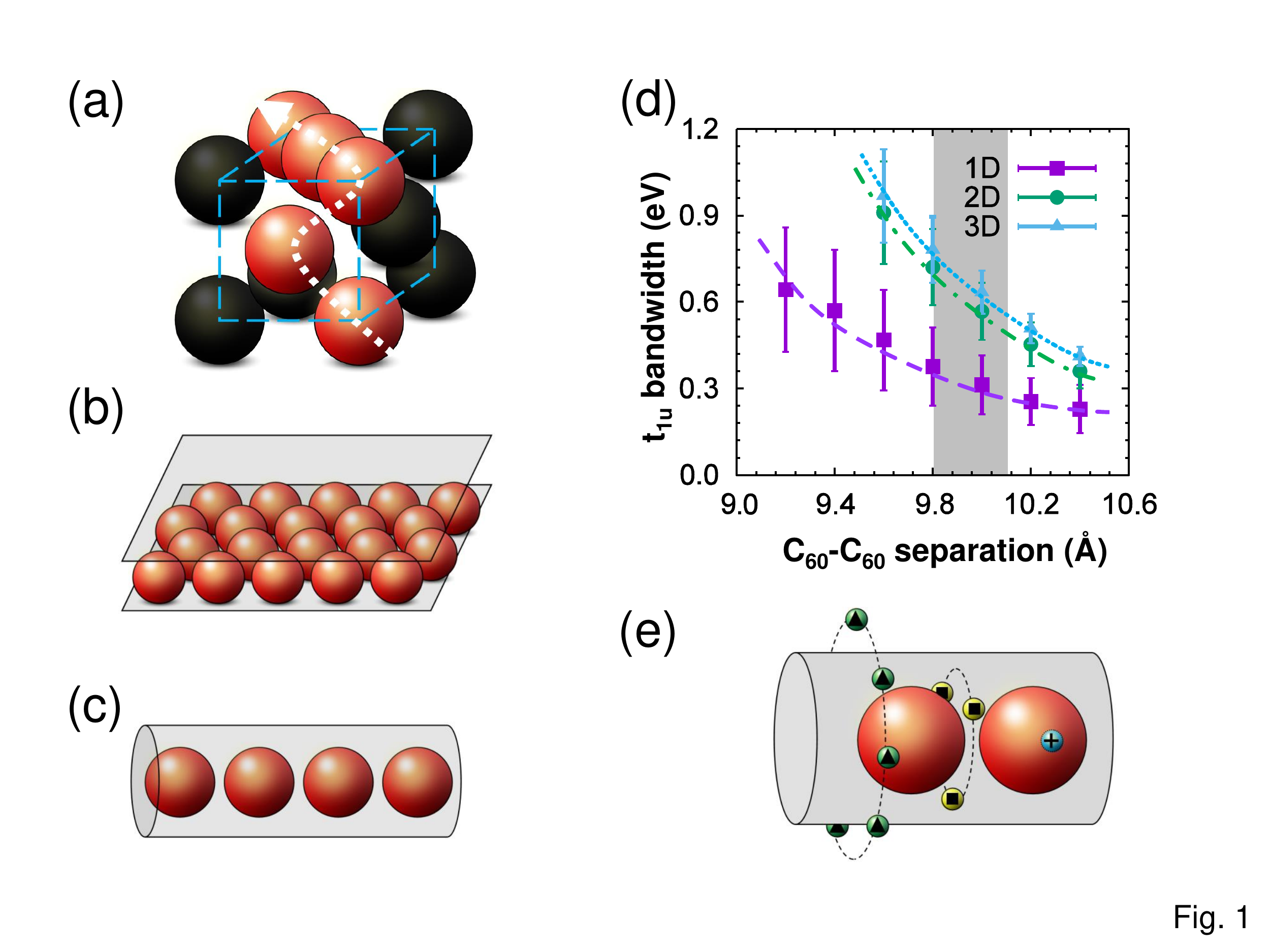}
\end{center}
\caption{
(Color online) Schematic arrangement of C$_{60}$ molecules in a
pristine (a) 3D fcc crystal, (b) 2D triangular lattice, and (c) 1D
array inside a $(10,10)$ carbon nanotube peapod. (d) Width of the
$t_{1u}$-derived band in a 1D, 2D and 3D arrangement of C$_{60}$
molecules as a function of the C$_{60}$-C$_{60}$ center-to-center
separation $d_{cc}$. %
Equilibrium values of $d_{cc}$
are affected by the C$_{60}$ orientational disorder, as 
indicated by the gray strip in (d) for undoped structures. The
`error bars' reflect the effect of changing the C$_{60}$
orientation on the bandwidth. (e) Schematic arrangement of dopant
atoms outside the 1D peapod ($\blacktriangle$), inside the
nanotube but outside the fullerene ($\scriptstyle\blacksquare$),
and inside the fullerene ({\bf +}). Dark spheres in (a) represent
clusters other than C$_{60}$ that separate quasi-1D percolating
arrays of fullerenes from the surrounding matrix. The planes in
(b) are only a visual aid. %
\label{fig1}}
\end{figure}

In alkali-doped M$_3$C$_{60}$ (M=K, Rb, Cs) fcc crystals,
superconductivity with $T_c{\lesssim}40$~K has been
observed~\cite{C60supHebard} and explained by electron-phonon
coupling that is modulated by the lattice
constant~\cite{{DT053},{DT057}}. The same behavior is expected to
occur in the isoelectronic La@C$_{60}$ that has been isolated from
raw soot~\cite{Shinohara-private} and found to be
stable~\cite{DT261}. When exohedrally doped M$_3$C$_{60}$ crystals
are exposed to ambient or harsh conditions, atoms from the
environment
may penetrate deep inside the lattice, react with the M atoms and
destroy superconductivity. This is much less likely to occur in
endohedrally doped La@C$_{60}$ crystals, since the dopant La atoms
are enclosed inside the protective C$_{60}$ cage. As mentioned
above, superconductivity in 3D M$_3$C$_{60}$ crystals is caused by
strong electron-phonon coupling related to a dynamical Jahn-Teller
effect on individual C$_{60}$ cages, made possible by retardation.
The dominant role of the intercalated alkali atoms is to partly
fill the $t_{1u}$ LUMO of C$_{60}$ that broadens to a narrow band
in the M$_3$C$_{60}$ molecular solid. Changes in $T_c$ caused by
pressure or changing the element M can be traced back to changes
in the electron-phonon coupling constant $\lambda=VN(E_F)$ in the
McMillan equation~\cite{{DT053},{DT057},{McMillan68}}. Since the
on-ball Bardeen-Pines interaction $V$ does not change, $\lambda$
is proportional to the C$_{60}$-projected DOS at the Fermi level
$N(E_F)$, which -- for electron doping -- is roughly inversely
proportional to the width of the $t_{1u}$-derived band. In
hole-doped C$_{60}$, $E_F$ is expected to be lowered into the
$h_u$-derived band with an even higher $N(E_F)$ value, which may
be the cause of the high value $T_c{\agt}60$~K that has been
reported earlier~\cite{C60SC70K}.

\begin{figure}[b]
\begin{center}
\includegraphics[width=0.92\columnwidth]{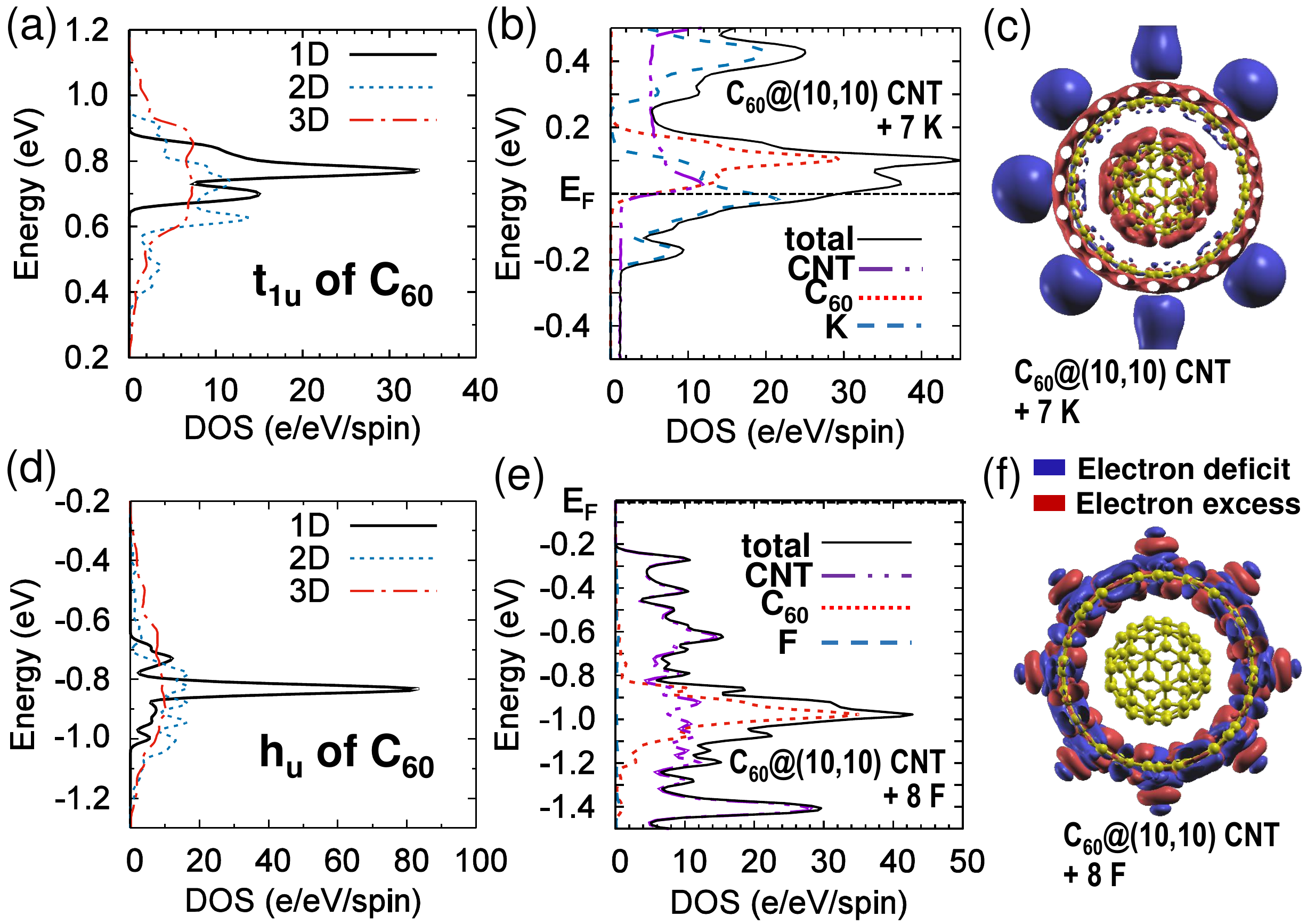}
\end{center}
\caption{%
(Color online) Density of states (DOS) and charge redistribution
in doped C$_{60}$ crystals. (a) DOS of the $t_{1u}$ LUMO-derived
bands of C$_{60}$ in 1D, 2D and 3D periodic C$_{60}$ arrangements
for the C$_{60}$-C$_{60}$ separation $d_{cc}=9.8$~{\AA}. (b) Total
and projected DOS of a C$_{60}@(10,10)$ CNT peapod doped
externally by 7~K donor atoms per C$_{60}$. (c) Charge density
difference
${\Delta}{\rho}={\rho}_{tot}$(C$_{60}@$CNT+7K)$-{\rho}_{tot}$(C$_{60}@$CNT)
$-\sum_{at}{\rho}_{tot}$(K atom). (d) DOS of the $h_{u}$
HOMO-derived bands of C$_{60}$ in 1D, 2D and 3D periodic C$_{60}$
arrangements for the C$_{60}$-C$_{60}$ separation
$d_{cc}=9.8$~{\AA}. (e) Total and projected DOS of a
C$_{60}@(10,10)$ CNT peapod doped externally by 8~F acceptor atoms
per C$_{60}$. (f) Counterpart of (c) for C$_{60}@$CNT+8F. In (c)
and (f), blue contours for electron deficit are shown for
${\Delta}{\rho}=-3.0{\times}10^{-3}$~e/bohr$^3$ and red contours
for electron excess are shown for
${\Delta}{\rho}=+3.0{\times}10^{-3}$~e/bohr$^3$.
All energies are with respect to $E_F$.%
\label{fig2}}
\end{figure}

All experimental strategies used so far to raise $T_c$ have been
based on increasing the C$_{60}$-C$_{60}$ separation $d_{cc}$ in a
3D fcc crystal, which would lower the width of the LUMO- and
HOMO-derived bands and thus increase $N(E_F)$ in doped crystals.
Our approach is quite different~\cite{Service2001}: we consider
increasing $N(E_F)$ by reducing the number of C$_{60}$ nearest
neighbors. As seen in Fig.~\ref{fig1}(a), this may be achieved
simply in a 3D crystal by mixing C$_{60}$ with clusters of similar
size that do not interact with C$_{60}$, such as BN fullerenes. In
this case, the lowered C$_{60}$ coordination number would decrease
the width of the $t_{1u}$ and $h_u$ derived bands and thus
increase $N(E_F)$ in doped crystals. Other C$_{60}$ arrangements
with a lower $Z$ include 2D arrays of C$_{60}$ that could possibly
be intercalated in graphite~\cite{{Saito1994},{Fuhrer1994}}, as
seen in Fig.~\ref{fig1}(b), or 1D arrays of C$_{60}$ in
C$_{60}@$CNT
peapods~\cite{{Okada2005},{Timoshevskii2009},{Koretsune2011}}
shown in in Fig.~\ref{fig1}(c). As seen in Fig.~\ref{fig1}(d), the
width of the $t_{1u}$-derived band decreases both with increasing
the C$_{60}$-C$_{60}$ separation and with the reduction of
dimensionality that translates to the reduction of $Z$,
with 1D arrangements appearing optimal. Since superconductivity is
suppressed in truly 1D systems according to the Mermin-Wagner
theorem~\cite{MWT}, we consider bundles of weakly interacting
peapods instead of isolated 1D peapods. As we will show in the
following, the main role of the nanotube in these systems is to
provide a suitable enclosure that aligns C$_{60}$ molecules and
protects them from the ambient. Due to their weak interaction,
bundles of nanotubes have a very similar DOS as isolated
nanotubes. Since the same applies to peapods, we will consider an
isolated peapod a valid representative of a peapod bundle from the
viewpoint of electronic structure.

Even in pristine systems with no intercalants, we found the
C$_{60}$-C$_{60}$ separation $d_{cc}$ to depend on the C$_{60}$
orientation and the dimensionality of the system. Different
fullerene orientations, each with a specific optimum $d_{cc}$
value, were found to be energetically degenerate within
${\alt}2$~meV/atom and separated by minute activation barriers. At
nonzero temperatures during synthesis and observation, the
fullerenes will explore these orientational degrees of freedom
causing orientational disorder and changing $d_{cc}$, as observed
in the 3D lattice~\cite{C60rot91}. The equilibrium value of
$d_{cc}$ decreases by ${\approx}0.1$~{\AA} and its orientational
dependence increases when reducing the dimensionality to 2D and
1D. Also in view of the soft C$_{60}$-C$_{60}$ interaction, we
always expect a nonzero range of $d_{cc}$ values in any
experimental sample. In pristine C$_{60}$, we expect $d_{cc}$
values roughly covering the $9.8-10.1$~{\AA} range shown by the
dark band in Fig.~\ref{fig1}(d).

As seen in Fig.~\ref{fig1}(e), the geometry is more complex in
alkali intercalated peapods, where intercalant atoms may occupy
sites outside the nanotube, inside the nanotube but outside
C$_{60}$, or inside the C$_{60}$ molecule such as the La@C$_{60}$
metallofullerene~\cite{DT261}. Since also these sites are
energetically near-degenerate, the precise geometry may be barely
controllable during synthesis. In the 3D M$_3$C$_{60}$ system,
moreover, $d_{cc}$ has been found to increase from 9.8~{\AA} to
10.3~{\AA} with increasing atomic number of the alkali element
M~\cite{DT053}. We find a similar M-dependent increase in $d_{cc}$
also in 2D and 1D systems, where M atoms separate fullerenes. 1D
peapods with the narrowest bandwidth and potentially highest
$N(E_F)$ could be doped by donor or acceptor atoms.
For most of this study, we will focus on donor doping, causing
partial filling of the $t_{1u}$-derived band, and will show later
that acceptor doping may be hard to achieve.

The DOS shape of the $t_{1u}$ LUMO-derived band in quasi-1D, 2D
triangular and 3D fcc lattices of C$_{60}$ is depicted in
Fig.~\ref{fig2}(a) and that of the $h_u$ HOMO-derived band in the
same lattices is shown in Fig.~\ref{fig2}(d). Clearly, the DOS at
$E_F$ reaches its maximum near half-filling of these bands in
quasi-1D structures. Since the $t_{1u}$ LUMO-derived band holds up
to 6 electrons and the $h_u$ HOMO-derived band up to 10 electrons,
half-filling of these bands requires either 3 extra electrons
or depletion of 5 electrons from each C$_{60}$.
Comparing our results in Figs.~\ref{fig2}(a) and \ref{fig2}(d), we
note that acceptor doping -- if achievable -- would result in a
significantly higher $N(E_F)$ than donor doping.

The calculated DOS of a C$_{60}@(10,10)$ peapod doped externally
by 7~K atoms per C$_{60}$ is shown in Fig.~\ref{fig2}(b) and the
DOS of the corresponding peapod doped externally by 8~F atoms per
C$_{60}$ is shown in Fig.~\ref{fig2}(e). Comparing the partial
densities of states in these two cases, we conclude that
C$_{60}$-derived states are barely affected by those of the
surrounding nanotube due to a very small hybridization. In the
case of donor doping by K depicted in Fig.~\ref{fig2}(b), we
clearly observe partial filling of the $t_{1u}$-derived band of
C$_{60}$ as well as the nearly-free electron bands of the
$(10,10)$ nanotube~\cite{Oshiyamapeapod01}. To get a better feel
for the charge flow in the system, we plotted the charge density
difference defined by
${\Delta}{\rho}={\rho}_{tot}$(C$_{60}@$CNT+7K)$-{\rho}_{tot}$(C$_{60}@$CNT)
$-\sum_{at}{\rho}_{tot}$(K atom) in Fig.~\ref{fig2}(c). Obviously,
there is a net electron flow from K atoms to the C$_{60}@(10,10)$
peapod, with the excess charge accommodated both by the C$_{60}$
and the nanotube. Integration of the C$_{60}$-projected DOS in
Fig.~\ref{fig2}(b) up to $E_F$ indicates a partial population of
the $t_{1u}$-derived band by $0.4$~electrons. %

The calculated DOS of an acceptor-doped peapod, shown in
Fig.~\ref{fig2}(e), presents a very different picture. We selected
F as a suitable electron acceptor due to its high
electronegativity. Unlike in previous studies of acceptor-doped
C$_{60}$, where covalently bonded halogen atoms disrupted the
$\pi$-electron network on the molecules~\cite{Saito92}, F atoms
were bonded on the outside of the nanotube surrounding C$_{60}$
molecules. Thus,
we found the $h_u$-derived band of C$_{60}$ to be essentially
unaffected by the presence of the surrounding nanotube and the 8~F
atoms per C$_{60}$ outside the nanotube, but the C$_{60}$
molecules remained charge neutral. The $h_u$-derived narrow band
band remained completely filled, located about 1~eV below $E_F$.
We found F atoms to bind covalently to the outside of the
nanotube, causing pyramidalization, disrupting its $\pi$-electron
network and opening up a gap at the Fermi level, which turned the
system into a semiconductor. This can be clearly seen when
inspecting the charge flow in this system in Fig.~\ref{fig2}(f).
We found F atoms to strongly hybridize with the C atoms of the
tube, redistributing the charge only within the F/CNT subsystem,
with no effect on the net charge of C$_{60}$. Since hole doping of
C$_{60}$ appears very difficult, we will focus on electron doping
of the $t_{1u}$-derived band of C$_{60}$ chains in the following.

\begin{figure}[tb]
\begin{center}
\includegraphics[width=0.92\columnwidth]{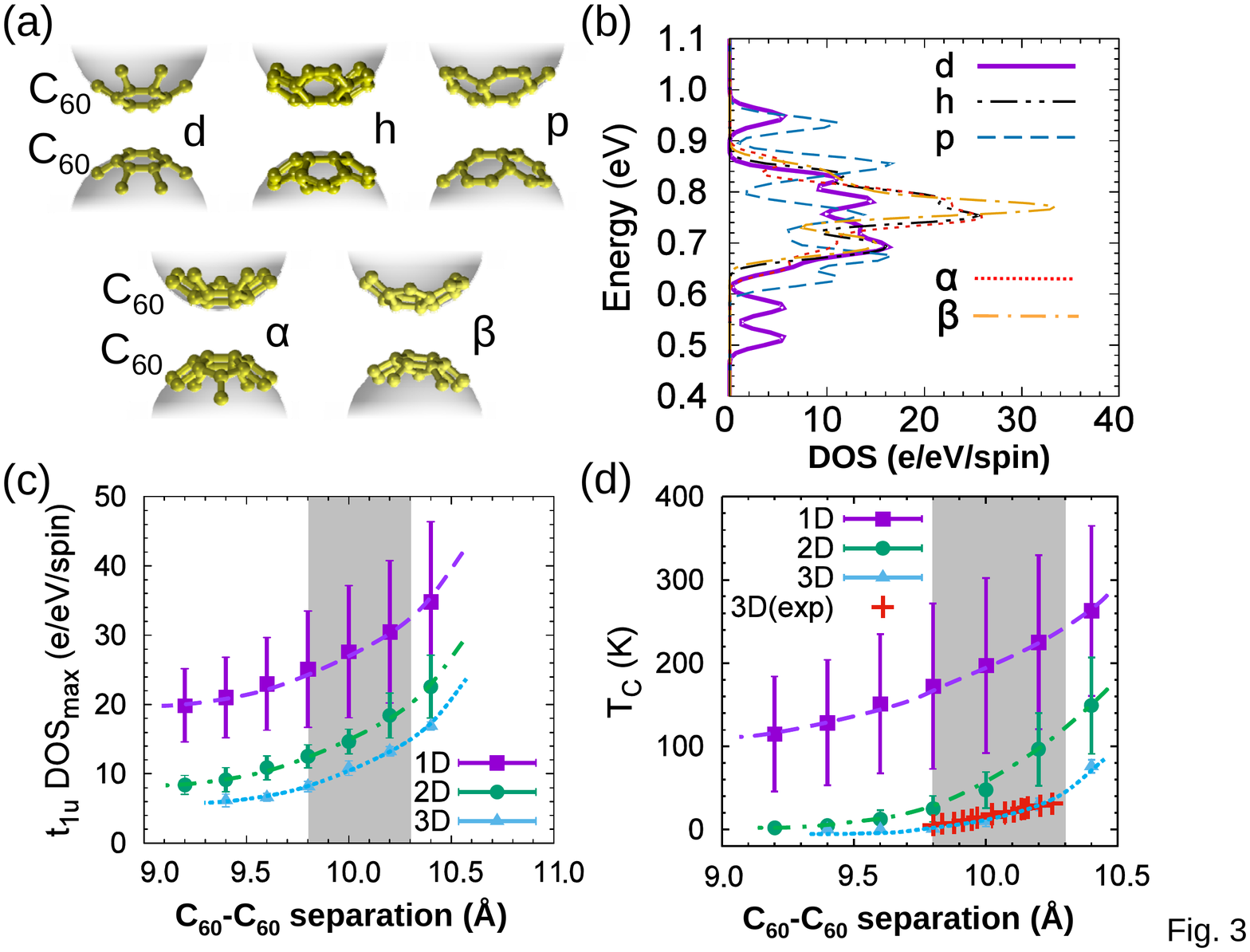}
\end{center}
\caption{%
(Color online) Properties of 1D arrays of C$_{60}$ molecules found
in CNT peapods. (a) Ball-and-stick models of different atomic
arrangements at the C$_{60}$-C$_{60}$ interface. Considered are
double-bonds facing double-bonds ({\em d}), hexagons facing
hexagons ({\em h}), pentagons facing pentagons ({\em p}). $\alpha$
and $\beta$ arrangements are obtained by a $90^\circ$ rotation of
C$_{60}$ molecules in {\em p} arrangement about two different axes
that are orthogonal to the chain axis and to each other. (b) DOS
of the $t_{1u}$ LUMO-derived bands of C$_{60}$ for the C$_{60}$
orientations, defined in (a), at the C$_{60}$-C$_{60}$ separation
$d_{cc}=9.8$~{\AA}. All energies are with respect to $E_F$. (c)
Maximum DOS value of $t_{1u}$-derived bands in 1D, 2D and 3D
crystals of C$_{60}$. The `error bars' reflect the effect of
changing the C$_{60}$ orientation. (d) Critical temperature for
superconductivity $T_c$ based on the McMillan equation
(\protect\ref{Eq1}) and using $N(E_F)$ from (c). Equilibrium
values of $d_{cc}$ %
are affected by the C$_{60}$ orientational disorder, as 
indicated by the gray strips in (c) and (d) for doped structures.
The lines in (c) and (d) are guides to the eye. %
\label{fig3}}
\end{figure}

As mentioned earlier, the electronic band structure of C$_{60}$
arrays should depend to a nontrivial degree on the orientation of
the C$_{60}$ molecules that will affect their
interaction~\cite{Gelfand92}.
We studied 5 different orientations, identified in
Fig.~\ref{fig3}(a), which result in a different degree of
inter-ball hybridization.
Due to their energetic near-degeneracy, we expect many C$_{60}$
orientations to coexist within a quasi-1D C$_{60}$ array inside a
peapod. The DOS for a chain of C$_{60}$ molecules at different
orientations and the C$_{60}$-C$_{60}$ separation
$d_{cc}=9.8$~{\AA} is shown in Fig.~\ref{fig3}(b). We note that
the maximum DOS value changes significantly with orientation.
Therefore, in Fig.~\ref{fig3}(c), we plotted the range of
achievable maxima of $N(E_F)$ as `error bars' for different
C$_{60}$-C$_{60}$ orientations.
%
Depending on the exact position of the intercalant atoms and
fullerene orientation, we found the optimum C$_{60}$-C$_{60}$
separations to cover the range $d_{cc}{\approx}9.8-10.3$~{\AA},
indicated by the gray strip in Fig.~\ref{fig3}(c). Higher $d_{cc}$
values than in pristine peapods, achieved in case that the
fullerenes are separated by heavy alkalis such as Cs, result in
very high values of $N(E_F)$ for favorable C$_{60}$ orientations.

To estimate the critical temperature for superconductivity $T_c$,
we used McMillan's equation~\cite{{DT053},{DT057},{McMillan68}} %
\begin{equation}
T_c = \frac{{\hbar}\omega_{log}}{1.2 k_B} %
\exp \left[ \frac{-1.04 (1+\lambda)} %
{\lambda-\mu^*-0.62\lambda\mu^*} \right] \,. %
\label{Eq1}
\end{equation}
This equation describes the solution of the Eliashberg equations
in superconductors with a strong electron-phonon coupling
($\lambda{\alt}2$) in a semi-empirical way that is physically
appealing. It also has been found to correctly reproduce the
observed $T_c$ values in M$_3$C$_{60}$ solids as a function of the
C$_{60}$-C$_{60}$ separation~\cite{DT057}, shown by the data
points for 3D systems in Fig.~\ref{fig3}(d). We used the
parameters of Ref.~[{\onlinecite{DT057}], namely
${\hbar}\omega_{log}/k_B=2800$~K, $\mu^*=0.2$ for the effective
mass and $V=52$~meV for the Bardeen-Pines interaction, which are
not affected by the averaged local arrangement of C$_{60}$
molecules. Using $\lambda=VN(E_F)$ for the electron-phonon
coupling constant, we were able to convert $N(E_F)$ values for 3D,
2D and quasi-1D systems with different C$_{60}$-C$_{60}$ distances
and C$_{60}$ orientations to potentially achievable $T_c$ values
and present our results in Fig.~\ref{fig3}(d).
Since dynamical orientational disorder and resulting fluctuations
in the C$_{60}$-C$_{60}$ distance are a natural phenomenon that is
particularly prominent in realistic 1D systems, we can estimate
ranges of $d_{cc}$ and $T_c$ values at best. 
Our estimates indicate that, in the best imaginable scenario,
$T_c$ near room temperature may be achievable using bundles of
donor-doped peapods.

Clearly, there are limits to the range of C$_{60}$-C$_{60}$
separations $d_{cc}$ compatible with superconductivity. Increasing
$d_{cc}$ decreases the inter-ball hopping integral $t$, while not
affecting the on-ball Coulomb integral $U$. At large
C$_{60}$-C$_{60}$ separations, the $U/t$ ratio should increase
beyond a critical value that would change doped C$_{60}$ from a
metal to a Mott-Hubbard
insulator~\cite{{Ganin-10},{Nomura-eC60-15}}.

In summary, we have proposed a viable way to further increase
$T_c$ for superconductivity by increasing the C$_{60}$-projected
density of states (DOS) at the Fermi level $N(E_F)$ and thus the
electron-phonon coupling constant in doped low-dimensional
C$_{60}$ nanoarrays. We considered intercalation by both electron
donors and acceptors, as well as electron doping in a solid formed
of La@C$_{60}$ endohedral complexes. We found that $N(E_F)$
increases with decreasing bandwidth of the partly filled $h_u$
HOMO- and $t_{1u}$ LUMO-derived frontier bands, which may be
achieved by reducing the coordination number of C$_{60}$. $N(E_F)$
increases significantly by changing from 3D C$_{60}$ crystals to
2D arrays of doped fullerenes intercalated in-between graphene
layers and reaches its maximum in doped quasi-1D arrays of
C$_{60}$ molecules inside C$_{60}$@CNT peapods formed of $(10,10)$
CNTs. Whereas partial filling of the $t_{1u}$-derived band may be
achieved by adsorbing alkali atoms outside the 1D peapod, the
desired depopulation of the $h_u$-derived band could not be
achieved by F atoms adsorbed on the nanotube surrounding the
C$_{60}$ molecules. Our results indicate that the highest $T_c$
values may occur in electron-doped C$_{60}$ peapods containing Cs
or in dilute 3D crystals, where quasi-1D arrangements of C$_{60}$
form percolation paths. Only experimental evidence will show if
low-dimensional arrays of doped C$_{60}$ will become
superconducting with $T_c$ approaching room temperature, or rather
turn to a Mott-Hubbard insulator.

\section*{Acknowledgments}

\begin{acknowledgments}
D.E. acknowledges the hospitality of MSU, where this research was
performed. D.L. and D.T. acknowledges financial support by the
NSF/AFOSR EFRI 2-DARE grant number EFMA-1433459. Computational
resources have been provided by the %
Michigan State University High Performance Computing Center.
\end{acknowledgments}



\begin{thebibliography}{29}%
\makeatletter
\providecommand \@ifxundefined [1]{%
 \@ifx{#1\undefined}
}%
\providecommand \@ifnum [1]{%
 \ifnum #1\expandafter \@firstoftwo
 \else \expandafter \@secondoftwo
 \fi
}%
\providecommand \@ifx [1]{%
 \ifx #1\expandafter \@firstoftwo
 \else \expandafter \@secondoftwo
 \fi
}%
\providecommand \natexlab [1]{#1}%
\providecommand \enquote  [1]{``#1''}%
\providecommand \bibnamefont  [1]{#1}%
\providecommand \bibfnamefont [1]{#1}%
\providecommand \citenamefont [1]{#1}%
\providecommand \href@noop [0]{\@secondoftwo}%
\providecommand \href [0]{\begingroup \@sanitize@url \@href}%
\providecommand \@href[1]{\@@startlink{#1}\@@href}%
\providecommand \@@href[1]{\endgroup#1\@@endlink}%
\providecommand \@sanitize@url [0]{\catcode `\\12\catcode
`\$12\catcode
  `\&12\catcode `\#12\catcode `\^12\catcode `\_12\catcode `\%12\relax}%
\providecommand \@@startlink[1]{}%
\providecommand \@@endlink[0]{}%
\providecommand \url  [0]{\begingroup\@sanitize@url \@url }%
\providecommand \@url [1]{\endgroup\@href {#1}{\urlprefix }}%
\providecommand \urlprefix  [0]{URL }%
\providecommand \Eprint [0]{\href }%
\providecommand \doibase [0]{http://dx.doi.org/}%
\providecommand \selectlanguage [0]{\@gobble}%
\providecommand \bibinfo  [0]{\@secondoftwo}%
\providecommand \bibfield  [0]{\@secondoftwo}%
\providecommand \translation [1]{[#1]}%
\providecommand \BibitemOpen [0]{}%
\providecommand \bibitemStop [0]{}%
\providecommand \bibitemNoStop [0]{.\EOS\space}%
\providecommand \EOS [0]{\spacefactor3000\relax}%
\providecommand \BibitemShut  [1]{\csname bibitem#1\endcsname}%
\let\auto@bib@innerbib\@empty
\bibitem [{\citenamefont {Bednorz}\ and\ \citenamefont
  {M\"{u}ller}(1986)}]{LBCO1986}%
  \BibitemOpen
  \bibfield  {author} {\bibinfo {author} {\bibfnamefont {J.~G.}\ \bibnamefont
  {Bednorz}}\ and\ \bibinfo {author} {\bibfnamefont {K.~A.}\ \bibnamefont
  {M\"{u}ller}},\ }\href {\doibase 10.1007/BF01303701} {\bibfield  {journal}
  {\bibinfo  {journal} {Z. Phys. B}\ }\textbf {\bibinfo {volume} {64}},\
  \bibinfo {pages} {189} (\bibinfo {year} {1986})}\BibitemShut {NoStop}%
\bibitem [{\citenamefont {Wu}\ \emph {et~al.}(1987)\citenamefont {Wu},
  \citenamefont {Ashburn}, \citenamefont {Torng}, \citenamefont {Hor},
  \citenamefont {Meng}, \citenamefont {Gao}, \citenamefont {Huang},
  \citenamefont {Wang},\ and\ \citenamefont {Chu}}]{YBCO1987}%
  \BibitemOpen
  \bibfield  {author} {\bibinfo {author} {\bibfnamefont {M.~K.}\ \bibnamefont
  {Wu}}, \bibinfo {author} {\bibfnamefont {J.~R.}\ \bibnamefont {Ashburn}},
  \bibinfo {author} {\bibfnamefont {C.~J.}\ \bibnamefont {Torng}}, \bibinfo
  {author} {\bibfnamefont {P.~H.}\ \bibnamefont {Hor}}, \bibinfo {author}
  {\bibfnamefont {R.~L.}\ \bibnamefont {Meng}}, \bibinfo {author}
  {\bibfnamefont {L.}~\bibnamefont {Gao}}, \bibinfo {author} {\bibfnamefont
  {Z.~J.}\ \bibnamefont {Huang}}, \bibinfo {author} {\bibfnamefont {Y.~Q.}\
  \bibnamefont {Wang}}, \ and\ \bibinfo {author} {\bibfnamefont {C.~W.}\
  \bibnamefont {Chu}},\ }\href {\doibase 10.1103/PhysRevLett.58.908} {\bibfield
   {journal} {\bibinfo  {journal} {Phys. Rev. Lett.}\ }\textbf {\bibinfo
  {volume} {58}},\ \bibinfo {pages} {908} (\bibinfo {year} {1987})}\BibitemShut
  {NoStop}%
\bibitem [{\citenamefont {Chu}\ \emph {et~al.}(1993)\citenamefont {Chu},
  \citenamefont {Gao}, \citenamefont {Chen}, \citenamefont {Huang},
  \citenamefont {Meng},\ and\ \citenamefont {Xue}}]{Highest-TC}%
  \BibitemOpen
  \bibfield  {author} {\bibinfo {author} {\bibfnamefont {C.~W.}\ \bibnamefont
  {Chu}}, \bibinfo {author} {\bibfnamefont {L.}~\bibnamefont {Gao}}, \bibinfo
  {author} {\bibfnamefont {F.}~\bibnamefont {Chen}}, \bibinfo {author}
  {\bibfnamefont {Z.~J.}\ \bibnamefont {Huang}}, \bibinfo {author}
  {\bibfnamefont {R.~L.}\ \bibnamefont {Meng}}, \ and\ \bibinfo {author}
  {\bibfnamefont {Y.~Y.}\ \bibnamefont {Xue}},\ }\href@noop {} {\bibfield
  {journal} {\bibinfo  {journal} {Nature}\ }\textbf {\bibinfo {volume} {365}},\
  \bibinfo {pages} {323} (\bibinfo {year} {1993})}\BibitemShut {NoStop}%
\bibitem [{\citenamefont {Drozdov}\ \emph {et~al.}(2015)\citenamefont
  {Drozdov}, \citenamefont {Eremets}, \citenamefont {Troyan}, \citenamefont
  {Ksenofontov},\ and\ \citenamefont {Shylin}}]{drozdov2015}%
  \BibitemOpen
  \bibfield  {author} {\bibinfo {author} {\bibfnamefont {A.}~\bibnamefont
  {Drozdov}}, \bibinfo {author} {\bibfnamefont {M.}~\bibnamefont {Eremets}},
  \bibinfo {author} {\bibfnamefont {I.}~\bibnamefont {Troyan}}, \bibinfo
  {author} {\bibfnamefont {V.}~\bibnamefont {Ksenofontov}}, \ and\ \bibinfo
  {author} {\bibfnamefont {S.}~\bibnamefont {Shylin}},\ }\href {\doibase
  10.1038/nature14964} {\bibfield  {journal} {\bibinfo  {journal} {Nature}\
  }\textbf {\bibinfo {volume} {525}},\ \bibinfo {pages} {73} (\bibinfo {year}
  {2015})}\BibitemShut {NoStop}%
\bibitem [{\citenamefont {Song}\ \emph {et~al.}(1993)\citenamefont {Song},
  \citenamefont {Fredette}, \citenamefont {Chung},\ and\ \citenamefont
  {Kao}}]{C60SC70K}%
  \BibitemOpen
  \bibfield  {author} {\bibinfo {author} {\bibfnamefont {L.}~\bibnamefont
  {Song}}, \bibinfo {author} {\bibfnamefont {K.}~\bibnamefont {Fredette}},
  \bibinfo {author} {\bibfnamefont {D.}~\bibnamefont {Chung}}, \ and\ \bibinfo
  {author} {\bibfnamefont {Y.}~\bibnamefont {Kao}},\ }\href {\doibase
  http://dx.doi.org/10.1016/0038-1098(93)90782-I} {\bibfield  {journal}
  {\bibinfo  {journal} {Solid State Commun.}\ }\textbf {\bibinfo {volume}
  {87}},\ \bibinfo {pages} {387} (\bibinfo {year} {1993})}\BibitemShut
  {NoStop}%
\bibitem [{\citenamefont {Schluter}\ \emph
  {et~al.}(1992{\natexlab{a}})\citenamefont {Schluter}, \citenamefont {Lannoo},
  \citenamefont {Needels}, \citenamefont {Baraff},\ and\ \citenamefont
  {Tomanek}}]{DT053}%
  \BibitemOpen
  \bibfield  {author} {\bibinfo {author} {\bibfnamefont {M.}~\bibnamefont
  {Schluter}}, \bibinfo {author} {\bibfnamefont {M.}~\bibnamefont {Lannoo}},
  \bibinfo {author} {\bibfnamefont {M.}~\bibnamefont {Needels}}, \bibinfo
  {author} {\bibfnamefont {G.~A.}\ \bibnamefont {Baraff}}, \ and\ \bibinfo
  {author} {\bibfnamefont {D.}~\bibnamefont {Tomanek}},\ }\href {\doibase
  10.1103/PhysRevLett.68.526} {\bibfield  {journal} {\bibinfo  {journal} {Phys.
  Rev. Lett.}\ }\textbf {\bibinfo {volume} {68}},\ \bibinfo {pages} {526}
  (\bibinfo {year} {1992}{\natexlab{a}})}\BibitemShut {NoStop}%
\bibitem [{\citenamefont {Schluter}\ \emph
  {et~al.}(1992{\natexlab{b}})\citenamefont {Schluter}, \citenamefont {Lannoo},
  \citenamefont {Needels}, \citenamefont {Baraff},\ and\ \citenamefont
  {Tomanek}}]{DT057}%
  \BibitemOpen
  \bibfield  {author} {\bibinfo {author} {\bibfnamefont {M.}~\bibnamefont
  {Schluter}}, \bibinfo {author} {\bibfnamefont {M.}~\bibnamefont {Lannoo}},
  \bibinfo {author} {\bibfnamefont {M.}~\bibnamefont {Needels}}, \bibinfo
  {author} {\bibfnamefont {G.~A.}\ \bibnamefont {Baraff}}, \ and\ \bibinfo
  {author} {\bibfnamefont {D.}~\bibnamefont {Tomanek}},\ }\href {\doibase
  10.1016/0022-3697(92)90240-E} {\bibfield  {journal} {\bibinfo  {journal} {J.
  Phys. Chem. Solids}\ }\textbf {\bibinfo {volume} {53}},\ \bibinfo {pages}
  {1473} (\bibinfo {year} {1992}{\natexlab{b}})}\BibitemShut {NoStop}%
\bibitem [{\citenamefont {McMillan}(1968)}]{McMillan68}%
  \BibitemOpen
  \bibfield  {author} {\bibinfo {author} {\bibfnamefont {W.~L.}\ \bibnamefont
  {McMillan}},\ }\href {\doibase 10.1103/PhysRev.167.331} {\bibfield  {journal}
  {\bibinfo  {journal} {Phys. Rev.}\ }\textbf {\bibinfo {volume} {167}},\
  \bibinfo {pages} {331} (\bibinfo {year} {1968})}\BibitemShut {NoStop}%
\bibitem [{\citenamefont {Perdew}\ and\ \citenamefont
  {Zunger}(1981)}]{Perdew81}%
  \BibitemOpen
  \bibfield  {author} {\bibinfo {author} {\bibfnamefont {J.~P.}\ \bibnamefont
  {Perdew}}\ and\ \bibinfo {author} {\bibfnamefont {A.}~\bibnamefont
  {Zunger}},\ }\href@noop {} {\bibfield  {journal} {\bibinfo  {journal} {Phys.
  Rev. B}\ }\textbf {\bibinfo {volume} {23}},\ \bibinfo {pages} {5048}
  (\bibinfo {year} {1981})}\BibitemShut {NoStop}%
\bibitem [{\citenamefont {Soler}\ \emph {et~al.}(2002)\citenamefont {Soler},
  \citenamefont {Artacho}, \citenamefont {Gale}, \citenamefont {Garc\'{i}a},
  \citenamefont {Junquera}, \citenamefont {Ordej\'{o}n},\ and\ \citenamefont
  {S\'{a}nchez-Portal}}]{soler-tsmfaioms2002}%
  \BibitemOpen
  \bibfield  {author} {\bibinfo {author} {\bibfnamefont {J.~M.}\ \bibnamefont
  {Soler}}, \bibinfo {author} {\bibfnamefont {E.}~\bibnamefont {Artacho}},
  \bibinfo {author} {\bibfnamefont {J.~D.}\ \bibnamefont {Gale}}, \bibinfo
  {author} {\bibfnamefont {A.}~\bibnamefont {Garc\'{i}a}}, \bibinfo {author}
  {\bibfnamefont {J.}~\bibnamefont {Junquera}}, \bibinfo {author}
  {\bibfnamefont {P.}~\bibnamefont {Ordej\'{o}n}}, \ and\ \bibinfo {author}
  {\bibfnamefont {D.}~\bibnamefont {S\'{a}nchez-Portal}},\ }\href@noop {}
  {\bibfield  {journal} {\bibinfo  {journal} {J. Phys: Cond. Mat.}\ }\textbf
  {\bibinfo {volume} {14}},\ \bibinfo {pages} {2745} (\bibinfo {year}
  {2002})}\BibitemShut {NoStop}%
\bibitem [{\citenamefont {Troullier}\ and\ \citenamefont
  {Martins}(1991)}]{Troullier91}%
  \BibitemOpen
  \bibfield  {author} {\bibinfo {author} {\bibfnamefont {N.}~\bibnamefont
  {Troullier}}\ and\ \bibinfo {author} {\bibfnamefont {J.~L.}\ \bibnamefont
  {Martins}},\ }\href@noop {} {\bibfield  {journal} {\bibinfo  {journal} {Phys.
  Rev. B}\ }\textbf {\bibinfo {volume} {43}},\ \bibinfo {pages} {1993}
  (\bibinfo {year} {1991})}\BibitemShut {NoStop}%
\bibitem [{\citenamefont {Kleinman}\ and\ \citenamefont
  {Bylander}(1982)}]{Kleinman82}%
  \BibitemOpen
  \bibfield  {author} {\bibinfo {author} {\bibfnamefont {L.}~\bibnamefont
  {Kleinman}}\ and\ \bibinfo {author} {\bibfnamefont {D.~M.}\ \bibnamefont
  {Bylander}},\ }\href@noop {} {\bibfield  {journal} {\bibinfo  {journal}
  {Phys. Rev. Lett.}\ }\textbf {\bibinfo {volume} {48}},\ \bibinfo {pages}
  {1425} (\bibinfo {year} {1982})}\BibitemShut {NoStop}%
\bibitem [{\citenamefont {Artacho}\ \emph {et~al.}(1999)\citenamefont
  {Artacho}, \citenamefont {S\'{a}nchez-Portal}, \citenamefont {Ordej\'{o}n},
  \citenamefont {Garc\'{\i}a},\ and\ \citenamefont {Soler}}]{SIESTA_PAO}%
  \BibitemOpen
  \bibfield  {author} {\bibinfo {author} {\bibfnamefont {E.}~\bibnamefont
  {Artacho}}, \bibinfo {author} {\bibfnamefont {D.}~\bibnamefont
  {S\'{a}nchez-Portal}}, \bibinfo {author} {\bibfnamefont {P.}~\bibnamefont
  {Ordej\'{o}n}}, \bibinfo {author} {\bibfnamefont {A.}~\bibnamefont
  {Garc\'{\i}a}}, \ and\ \bibinfo {author} {\bibfnamefont {J.~M.}\ \bibnamefont
  {Soler}},\ }\href@noop {} {\bibfield  {journal} {\bibinfo  {journal} {Phys.
  Stat. Sol.}\ }\textbf {\bibinfo {volume} {215}},\ \bibinfo {pages} {809}
  (\bibinfo {year} {1999})}\BibitemShut {NoStop}%
\bibitem [{\citenamefont {Hebard}\ \emph {et~al.}(1991)\citenamefont {Hebard},
  \citenamefont {Rosseinsky}, \citenamefont {Haddon}, \citenamefont {Murphy},
  \citenamefont {Glarum}, \citenamefont {Palstra}, \citenamefont {Ramirez},\
  and\ \citenamefont {Kortan}}]{C60supHebard}%
  \BibitemOpen
  \bibfield  {author} {\bibinfo {author} {\bibfnamefont {A.~F.}\ \bibnamefont
  {Hebard}}, \bibinfo {author} {\bibfnamefont {M.~J.}\ \bibnamefont
  {Rosseinsky}}, \bibinfo {author} {\bibfnamefont {R.~C.}\ \bibnamefont
  {Haddon}}, \bibinfo {author} {\bibfnamefont {D.~W.}\ \bibnamefont {Murphy}},
  \bibinfo {author} {\bibfnamefont {S.~H.}\ \bibnamefont {Glarum}}, \bibinfo
  {author} {\bibfnamefont {T.~T.~M.}\ \bibnamefont {Palstra}}, \bibinfo
  {author} {\bibfnamefont {A.~P.}\ \bibnamefont {Ramirez}}, \ and\ \bibinfo
  {author} {\bibfnamefont {A.~R.}\ \bibnamefont {Kortan}},\ }\href {\doibase
  10.1038/350600a0} {\bibfield  {journal} {\bibinfo  {journal} {Nature}\
  }\textbf {\bibinfo {volume} {350}},\ \bibinfo {pages} {600} (\bibinfo {year}
  {1991})}\BibitemShut {NoStop}%
\bibitem [{Shi()}]{Shinohara-private}%
  \BibitemOpen
  \href@noop {} {}\bibinfo {note} {H. Shinohara (private
  communication).}\BibitemShut {Stop}%
\bibitem [{\citenamefont {Guan}\ and\ \citenamefont {Tom\'anek}(2017)}]{DT261}%
  \BibitemOpen
  \bibfield  {author} {\bibinfo {author} {\bibfnamefont {J.}~\bibnamefont
  {Guan}}\ and\ \bibinfo {author} {\bibfnamefont {D.}~\bibnamefont
  {Tom\'anek}},\ }\href {\doibase 10.1021/acs.nanolett.7b00185} {\bibfield
  {journal} {\bibinfo  {journal} {Nano Lett.}\ }\textbf {\bibinfo {volume}
  {17}},\ \bibinfo {pages} {3402} (\bibinfo {year} {2017})}\BibitemShut
  {NoStop}%
\bibitem [{\citenamefont {Service}(2001)}]{Service2001}%
  \BibitemOpen
  \bibfield  {author} {\bibinfo {author} {\bibfnamefont {R.~F.}\ \bibnamefont
  {Service}},\ }\href {\doibase 10.1126/science.292.5514.45} {\bibfield
  {journal} {\bibinfo  {journal} {Science}\ }\textbf {\bibinfo {volume}
  {292}},\ \bibinfo {pages} {45} (\bibinfo {year} {2001})}\BibitemShut
  {NoStop}%
\bibitem [{\citenamefont {Saito}\ and\ \citenamefont
  {Oshiyama}(1994)}]{Saito1994}%
  \BibitemOpen
  \bibfield  {author} {\bibinfo {author} {\bibfnamefont {S.}~\bibnamefont
  {Saito}}\ and\ \bibinfo {author} {\bibfnamefont {A.}~\bibnamefont
  {Oshiyama}},\ }\href {\doibase 10.1103/PhysRevB.49.17413} {\bibfield
  {journal} {\bibinfo  {journal} {Phys. Rev. B}\ }\textbf {\bibinfo {volume}
  {49}},\ \bibinfo {pages} {17413} (\bibinfo {year} {1994})}\BibitemShut
  {NoStop}%
\bibitem [{\citenamefont {Fuhrer}\ \emph {et~al.}(1994)\citenamefont {Fuhrer},
  \citenamefont {Hou}, \citenamefont {Xiang},\ and\ \citenamefont
  {Zettl}}]{Fuhrer1994}%
  \BibitemOpen
  \bibfield  {author} {\bibinfo {author} {\bibfnamefont {M.}~\bibnamefont
  {Fuhrer}}, \bibinfo {author} {\bibfnamefont {J.}~\bibnamefont {Hou}},
  \bibinfo {author} {\bibfnamefont {X.-D.}\ \bibnamefont {Xiang}}, \ and\
  \bibinfo {author} {\bibfnamefont {A.}~\bibnamefont {Zettl}},\ }\href
  {\doibase 10.1016/0038-1098(94)90798-6} {\bibfield  {journal} {\bibinfo
  {journal} {Solid State Commun.}\ }\textbf {\bibinfo {volume} {90}},\ \bibinfo
  {pages} {357} (\bibinfo {year} {1994})}\BibitemShut {NoStop}%
\bibitem [{\citenamefont {Okada}(2005)}]{Okada2005}%
  \BibitemOpen
  \bibfield  {author} {\bibinfo {author} {\bibfnamefont {S.}~\bibnamefont
  {Okada}},\ }\href {\doibase 10.1103/PhysRevB.72.153409} {\bibfield  {journal}
  {\bibinfo  {journal} {Phys. Rev. B}\ }\textbf {\bibinfo {volume} {72}},\
  \bibinfo {pages} {153409} (\bibinfo {year} {2005})}\BibitemShut {NoStop}%
\bibitem [{\citenamefont {Timoshevskii}\ and\ \citenamefont
  {C\^ot\'e}(2009)}]{Timoshevskii2009}%
  \BibitemOpen
  \bibfield  {author} {\bibinfo {author} {\bibfnamefont {V.}~\bibnamefont
  {Timoshevskii}}\ and\ \bibinfo {author} {\bibfnamefont {M.}~\bibnamefont
  {C\^ot\'e}},\ }\href {\doibase 10.1103/PhysRevB.80.235418} {\bibfield
  {journal} {\bibinfo  {journal} {Phys. Rev. B}\ }\textbf {\bibinfo {volume}
  {80}},\ \bibinfo {pages} {235418} (\bibinfo {year} {2009})}\BibitemShut
  {NoStop}%
\bibitem [{\citenamefont {Koretsune}\ \emph {et~al.}(2011)\citenamefont
  {Koretsune}, \citenamefont {Saito},\ and\ \citenamefont
  {Cohen}}]{Koretsune2011}%
  \BibitemOpen
  \bibfield  {author} {\bibinfo {author} {\bibfnamefont {T.}~\bibnamefont
  {Koretsune}}, \bibinfo {author} {\bibfnamefont {S.}~\bibnamefont {Saito}}, \
  and\ \bibinfo {author} {\bibfnamefont {M.~L.}\ \bibnamefont {Cohen}},\ }\href
  {\doibase 10.1103/PhysRevB.83.193406} {\bibfield  {journal} {\bibinfo
  {journal} {Phys. Rev. B}\ }\textbf {\bibinfo {volume} {83}},\ \bibinfo
  {pages} {193406} (\bibinfo {year} {2011})}\BibitemShut {NoStop}%
\bibitem [{\citenamefont {Mermin}\ and\ \citenamefont {Wagner}(1966)}]{MWT}%
  \BibitemOpen
  \bibfield  {author} {\bibinfo {author} {\bibfnamefont {N.~D.}\ \bibnamefont
  {Mermin}}\ and\ \bibinfo {author} {\bibfnamefont {H.}~\bibnamefont
  {Wagner}},\ }\href@noop {} {\bibfield  {journal} {\bibinfo  {journal} {Phys.
  Rev. Lett.}\ }\textbf {\bibinfo {volume} {17}},\ \bibinfo {pages} {1133}
  (\bibinfo {year} {1966})},\ \bibinfo {note} {and {\em ibid.} {\bf 17}, 1307
  (1966)(E)}\BibitemShut {NoStop}%
\bibitem [{\citenamefont {Heiney}\ \emph {et~al.}(1991)\citenamefont {Heiney},
  \citenamefont {Fischer}, \citenamefont {McGhie}, \citenamefont {Romanow},
  \citenamefont {Denenstein}, \citenamefont {McCauley~Jr.}, \citenamefont
  {Smith},\ and\ \citenamefont {Cox}}]{C60rot91}%
  \BibitemOpen
  \bibfield  {author} {\bibinfo {author} {\bibfnamefont {P.~A.}\ \bibnamefont
  {Heiney}}, \bibinfo {author} {\bibfnamefont {J.~E.}\ \bibnamefont {Fischer}},
  \bibinfo {author} {\bibfnamefont {A.~R.}\ \bibnamefont {McGhie}}, \bibinfo
  {author} {\bibfnamefont {W.~J.}\ \bibnamefont {Romanow}}, \bibinfo {author}
  {\bibfnamefont {A.~M.}\ \bibnamefont {Denenstein}}, \bibinfo {author}
  {\bibfnamefont {J.~P.}\ \bibnamefont {McCauley~Jr.}}, \bibinfo {author}
  {\bibfnamefont {A.~B.}\ \bibnamefont {Smith}}, \ and\ \bibinfo {author}
  {\bibfnamefont {D.~E.}\ \bibnamefont {Cox}},\ }\href {\doibase
  10.1103/PhysRevLett.66.2911} {\bibfield  {journal} {\bibinfo  {journal}
  {Phys. Rev. Lett.}\ }\textbf {\bibinfo {volume} {66}},\ \bibinfo {pages}
  {2911} (\bibinfo {year} {1991})}\BibitemShut {NoStop}%
\bibitem [{\citenamefont {Okada}\ \emph {et~al.}(2001)\citenamefont {Okada},
  \citenamefont {Saito},\ and\ \citenamefont {Oshiyama}}]{Oshiyamapeapod01}%
  \BibitemOpen
  \bibfield  {author} {\bibinfo {author} {\bibfnamefont {S.}~\bibnamefont
  {Okada}}, \bibinfo {author} {\bibfnamefont {S.}~\bibnamefont {Saito}}, \ and\
  \bibinfo {author} {\bibfnamefont {A.}~\bibnamefont {Oshiyama}},\ }\href
  {\doibase 10.1103/PhysRevLett.86.3835} {\bibfield  {journal} {\bibinfo
  {journal} {Phys. Rev. Lett.}\ }\textbf {\bibinfo {volume} {86}},\ \bibinfo
  {pages} {3835} (\bibinfo {year} {2001})}\BibitemShut {NoStop}%
\bibitem [{\citenamefont {Miyamoto}\ \emph {et~al.}(1992)\citenamefont
  {Miyamoto}, \citenamefont {Oshiyama},\ and\ \citenamefont {Saito}}]{Saito92}%
  \BibitemOpen
  \bibfield  {author} {\bibinfo {author} {\bibfnamefont {Y.}~\bibnamefont
  {Miyamoto}}, \bibinfo {author} {\bibfnamefont {A.}~\bibnamefont {Oshiyama}},
  \ and\ \bibinfo {author} {\bibfnamefont {S.}~\bibnamefont {Saito}},\ }\href
  {\doibase https://doi.org/10.1016/0038-1098(92)90745-U} {\bibfield  {journal}
  {\bibinfo  {journal} {Solid State Commun.}\ }\textbf {\bibinfo {volume}
  {82}},\ \bibinfo {pages} {437} (\bibinfo {year} {1992})}\BibitemShut
  {NoStop}%
\bibitem [{\citenamefont {Gelfand}\ and\ \citenamefont {Lu}(1992)}]{Gelfand92}%
  \BibitemOpen
  \bibfield  {author} {\bibinfo {author} {\bibfnamefont {M.~P.}\ \bibnamefont
  {Gelfand}}\ and\ \bibinfo {author} {\bibfnamefont {J.~P.}\ \bibnamefont
  {Lu}},\ }\href {\doibase 10.1103/PhysRevLett.68.1050} {\bibfield  {journal}
  {\bibinfo  {journal} {Phys. Rev. Lett.}\ }\textbf {\bibinfo {volume} {68}},\
  \bibinfo {pages} {1050} (\bibinfo {year} {1992})}\BibitemShut {NoStop}%
\bibitem [{\citenamefont {Ganin}\ \emph {et~al.}(2010)\citenamefont {Ganin},
  \citenamefont {Takabayashi}, \citenamefont {Jegli\v{c}}, \citenamefont
  {Ar\v{c}on}, \citenamefont {Poto\v{c}nik}, \citenamefont {Baker},
  \citenamefont {Ohishi}, \citenamefont {McDonald}, \citenamefont {Tzirakis},
  \citenamefont {McLennan}, \citenamefont {Darling}, \citenamefont {Takata},
  \citenamefont {Rosseinsky},\ and\ \citenamefont {Prassides}}]{Ganin-10}%
  \BibitemOpen
  \bibfield  {author} {\bibinfo {author} {\bibfnamefont {A.~Y.}\ \bibnamefont
  {Ganin}}, \bibinfo {author} {\bibfnamefont {Y.}~\bibnamefont {Takabayashi}},
  \bibinfo {author} {\bibfnamefont {P.}~\bibnamefont {Jegli\v{c}}}, \bibinfo
  {author} {\bibfnamefont {D.}~\bibnamefont {Ar\v{c}on}}, \bibinfo {author}
  {\bibfnamefont {A.}~\bibnamefont {Poto\v{c}nik}}, \bibinfo {author}
  {\bibfnamefont {P.~J.}\ \bibnamefont {Baker}}, \bibinfo {author}
  {\bibfnamefont {Y.}~\bibnamefont {Ohishi}}, \bibinfo {author} {\bibfnamefont
  {M.~T.}\ \bibnamefont {McDonald}}, \bibinfo {author} {\bibfnamefont {M.~D.}\
  \bibnamefont {Tzirakis}}, \bibinfo {author} {\bibfnamefont {A.}~\bibnamefont
  {McLennan}}, \bibinfo {author} {\bibfnamefont {G.~R.}\ \bibnamefont
  {Darling}}, \bibinfo {author} {\bibfnamefont {M.}~\bibnamefont {Takata}},
  \bibinfo {author} {\bibfnamefont {M.~J.}\ \bibnamefont {Rosseinsky}}, \ and\
  \bibinfo {author} {\bibfnamefont {K.}~\bibnamefont {Prassides}},\ }\href
  {\doibase 10.1038/nature09120} {\bibfield  {journal} {\bibinfo  {journal}
  {Nature}\ }\textbf {\bibinfo {volume} {466}},\ \bibinfo {pages} {221}
  (\bibinfo {year} {2010})}\BibitemShut {NoStop}%
\bibitem [{\citenamefont {Nomura}\ \emph {et~al.}(2015)\citenamefont {Nomura},
  \citenamefont {Sakai}, \citenamefont {Capone},\ and\ \citenamefont
  {Arita}}]{Nomura-eC60-15}%
  \BibitemOpen
  \bibfield  {author} {\bibinfo {author} {\bibfnamefont {Y.}~\bibnamefont
  {Nomura}}, \bibinfo {author} {\bibfnamefont {S.}~\bibnamefont {Sakai}},
  \bibinfo {author} {\bibfnamefont {M.}~\bibnamefont {Capone}}, \ and\ \bibinfo
  {author} {\bibfnamefont {R.}~\bibnamefont {Arita}},\ }\href {\doibase
  10.1126/sciadv.1500568} {\bibfield  {journal} {\bibinfo  {journal} {Science
  Adv.}\ }\textbf {\bibinfo {volume} {1}},\ \bibinfo {pages} {e1500568}
  (\bibinfo {year} {2015})}\BibitemShut {NoStop}%
\end{thebibliography}

%

\end{document}